\documentclass[aip, apl, reprint, citeautoscript]{revtex4-1}
\usepackage{graphicx}
\usepackage{amsmath}
\usepackage{amssymb}
\usepackage{xcolor}
\usepackage{ulem}

\begin{document}


\title{Atomic-Scale Tailoring of Chemisorbed Atomic Oxygen on Epitaxial Graphene for Graphene-Based Electronic Devices}



\author{Tae Soo Kim}
\thanks{T. S. Kim and T. Ahn contributed equally to this work.}
\affiliation{Department of Chemistry, Pohang University of Science and Technology (POSTECH), Pohang 37673, South Korea}
\affiliation{Center for Artificial Low Dimensional Electronic Systems, Institute for Basic Science (IBS), Pohang 37673, South Korea}

\author{Taemin Ahn}
\thanks{T. S. Kim and T. Ahn contributed equally to this work.}
\affiliation{Department of Physics, POSTECH, Pohang 37673, South Korea}

\author{Tae-Hwan Kim}
\email{taehwan@postech.ac.kr}
\affiliation{Department of Physics, POSTECH, Pohang 37673, South Korea}

\author{Hee Cheul Choi}
\affiliation{Department of Chemistry, Pohang University of Science and Technology (POSTECH), Pohang 37673, South Korea}
\affiliation{Center for Artificial Low Dimensional Electronic Systems, Institute for Basic Science (IBS), Pohang 37673, South Korea}

\author{Han Woong Yeom}
\affiliation{Center for Artificial Low Dimensional Electronic Systems, Institute for Basic Science (IBS), Pohang 37673, South Korea}
\affiliation{Department of Physics, POSTECH, Pohang 37673, South Korea}

\date{\today}

\begin{abstract}
Graphene, with its unique band structure, mechanical stability, and high charge mobility, holds great promise for next-generation electronics. 
Nevertheless, its zero band gap challenges the control of current flow through electrical gating, consequently limiting its practical applications. 
Recent research indicates that atomic oxygen can oxidize epitaxial graphene in a vacuum without causing unwanted damage. 
In this study, we have investigated the effects of chemisorbed atomic oxygen on the electronic properties of epitaxial graphene, using scanning tunneling microscopy (STM). 
Our findings reveal that oxygen atoms effectively modify the electronic states of graphene, resulting in a band gap at its Dirac point. 
Furthermore, we demonstrate that it is possible to selectively induce desorption or hopping of oxygen atoms with atomic precision by applying appropriate bias sweeps with an STM tip. 
These results suggest the potential for atomic-scale tailoring of graphene oxide, enabling the development of graphene-based atomic-scale electronic devices.
\end{abstract}


\maketitle 

Graphene, consisting of a single sheet of $sp^2$-bonded carbon atoms, presents a promising material candidate for future electronic devices due to its superior electron mobility~\cite{Novoselov2004a,Bolotin2008a,Morozov2008a} and mechanical stability~\cite{Lee2008a}. 
However, to harness its potential for practical applications, tuning its electronic structure and introducing various functionalizations are necessary. 
Oxidation is a well-established method for modifying the electronic~\cite{Eda2009a,Lopez2009a,Wei2012,Lipatov2018a} and optical~\cite{Luo2009a,Mathkar2012,Wei2012} properties of graphene. 
Nevertheless, most studies on graphene oxide have focused on samples prepared by oxidizing graphite powders in solution~\cite{Hummers1958a,Byon2011a}. 
As a result, the atomic-scale structures and electronic states of oxidized graphene largely remain unexplored~\cite{Pei2012a}. 
This ambiguity presents a significant limitation because the electronic properties of oxidized graphene are strongly dependent on functional groups, \textit{i.e.}, oxidation species. 
Therefore, it is highly desirable to characterize different oxidation species and their distribution at atomic resolution. 
Scanning tunneling microscopy (STM) is a powerful tool for this purpose~\cite{Rutter2007a,Stolyarova2007a,Zhang2009a,Kim2013a}, but it is technically challenging to apply STM measurements to samples prepared through the powder solution process~\cite{Pandey2008a}.

Recent developments in STM studies of graphene oxide have circumvented the difficulty of preparing samples through wet processes.
This is achieved by employing an \textit{in situ} dry oxidation process in a vacuum, 
which eliminates the risk of unwanted contamination typically introduced during wet processes.
This method has been  demonstrated on graphene grown on Pt(111) and Ir(111) surfaces~\cite{Vinogradov2011a}. 
Another STM experiment on epitaxial graphene grown on silicon carbide (SiC) has provided further atomistic details of the oxidation process, including the adsorption site of atomic oxygen and the recovery of the graphene lattice through thermal reduction~\cite{Hossain2012a}. 
In addition, spatially resolved scanning tunneling spectroscopy (STS) has revealed local changes in the electronic properties of monolayer graphene grown on copper foil due to the presence of oxygen atoms~\cite{Harthcock2017}.

In this letter, we present detailed STM and STS measurements of oxygen adsorption on epitaxial graphene. 
Our study  characterizes the electronic state of graphene in the presence of atomic oxygen adsorbates. 
Furthermore, we demonstrate that the STM tip can control the local removal or rearrangement of atomic oxygen by applying an appropriate bias sweep. 
This technique parallels previously reported STM nanolithography for hydrogenated Si(001)~\cite{Shen1995a,Fuechsle2012a,Bianco2013a,Achal2018} and graphene~\cite{Tapaszto2008a,Hossain2010a,Martinez-Galera2014,Diez-Albar2019}. 
The ability to control the desorption and hopping of chemisorbed oxygen atoms may allow us to tailor the local electronic properties of graphene with atomic precision.
This development holds significant potential for the creation of advanced graphene-based atomic-scale devices~\cite{Heinrich2002a}.

In this work, epitaxial graphene was grown on a silicon-terminated 6H- or 4H-SiC(0001) substrate ($n$-type, Cree, Inc.). 
The SiC substrates were first cleaned by sonication in isopropyl alcohol and acetone, then etched under a hydrogen flow at 1700~K to remove any mechanical polishing damage~\cite{Ostler2010,Yazdi2016}. 
A nominal thickness of 1.2~layers of epitaxial graphene was grown on the H$_2$-etched SiC substrate by annealing it at 1800~K in an Ar atmosphere~\cite{Ohta2006a,Virojanadara2008}.
After verifying the cleanliness of the epitaxial graphene surface on SiC through STM measurements, we proceeded to oxidize the epitaxial graphene by exposing it to atomic oxygen.
This atomic oxygen was generated by a home-built thermal cracker, which thermally decomposed O$_2$ using a hot (up to 1800~K) Ir capillary in the UHV chamber.
Thanks to the thermal cracker, we maintained an O$_2$ pressure as low as $\sim$3.0 $\times$ 10$^{-8}$~Torr during atomic oxygen exposure (typically 60~sec, corresponding to 1.8~L).
We conducted the STM/STS measurements at 80~K using a home-built low-temperature STM~\cite{our_stm} under a base pressure of less than $1.0\times10^{-10}$~Torr. 
All bias voltages refer to the sample voltage with respect to the STM tip.

\begin{figure} 
\includegraphics[width=8cm]{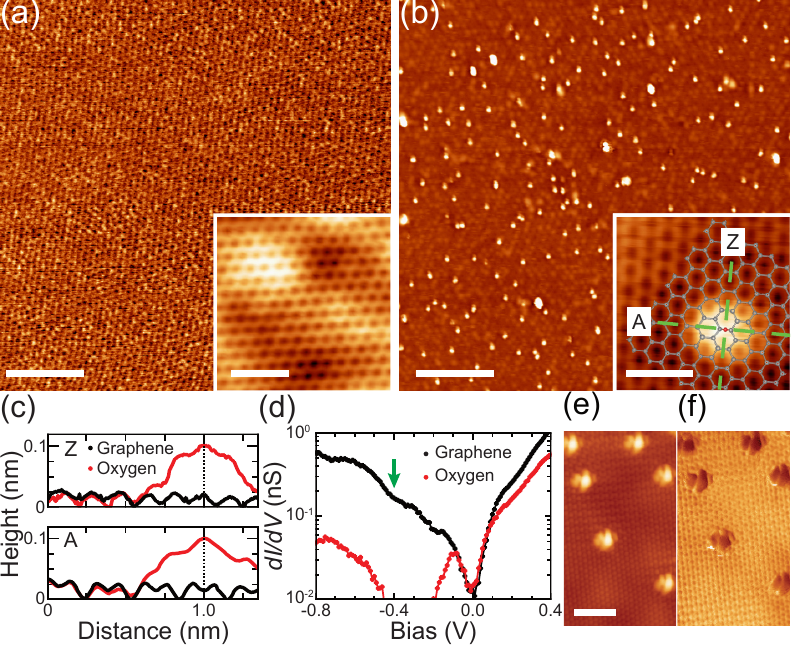}
\caption{\label{figure1}
(a) STM image of pristine epitaxial graphene on SiC(0001) before atomic oxygen exposure ($V_{\rm s}$ = $-$2.0~V, $I_{\rm t}$ = 10~pA, scale bar: 20~nm). 
The inset shows atom-resolved STM image of the graphene monolayer ($V_{\rm s}$ = 0.5 V, $I_{\rm t}$ = 50~pA, scale bar: 1~nm). 
(b) STM image of chemisorbed atomic oxygen on epitaxial graphene after atomic oxygen exposure up to a nominal 1.8~L ($V_{\rm s}$ = 2.0~V, $I_{\rm t}$ = 20~pA, scale bar: 20~nm). 
The inset exhibits an atom-resolved STM image of an isolated chemisorbed oxygen atom together with superimposed hexagonal graphene lattice ($V_{\rm s}$ = 2.0~V, $I_{\rm t}$ = 50~pA, scale bar: 1~nm). 
(c) Line profiles of the oxygen atom (red) and graphene lattice (black), measured along the zigzag (Z) and armchair (A) directions, as indicated by dashed lines in the inset of (b).
(d) Point STS spectra obtained on chemisorbed atomic oxygen (red) and graphene monolayer (black). 
The green arrow marks the Dirac point of the graphene monolayer (set point: $V_{\rm s}$ = 1.0~V, $I_{\rm t}$ = 50~pA, $Z_{\rm offset}$ = $-150$~pm)~\cite{sts}. 
(e) Topographic and (f) differential conductance images of chemisorbed atomic oxygen on graphene bilayer obtained at the Dirac point ($V_{\rm s}$ = $-0.3$~V), demonstrating the $dI/dV$ suppression at the Dirac point on chemisorbed atomic oxygen. 
The scale bar is 2~nm.
}
\end{figure}

Figure~1 shows STM images of epitaxial graphene grown on SiC both before [Fig.~1(a)] and after [Fig.~1(b)] exposure to atomic oxygen.
As shown in the atom-resolved STM image [inset of Fig.~1(b)], the graphene lattice and the oxygen protrusion can be observed simultaneously.
By analyzing the line profiles of the oxygen atom and the surrounding graphene lattice measured along the zigzag and armchair directions [Fig.~\ref{figure1}(c)], we determined that the oxygen atom's location on the epitaxial graphene corresponds to the epoxy binding site, which bridges the two nearest carbon atoms~\cite{Hossain2012a}.

Figure~1(d) exhibits STS spectra obtained on chemisorbed atomic oxygen (red) and the graphene monolayer (black). 
The spectrum for graphene identifies the Dirac point at approximately $-0.4$~V (marked by the green arrow) below the Fermi level, which aligns well with previous observations for epitaxial graphene monolayers~\cite{Lauffer2008a,Premlal2009a,Choi2010a}. 
In contrast, the spectrum derived from the oxygen protrusion shows a depression in the density of state (DOS) around the Dirac point.
We interpret this DOS depression as a band gap of about 250~meV, which is comparable to those reported in other studies~\cite{Harthcock2017}.

Moreover, the DOS suppression caused by chemisorbed atomic oxygen is vividly illustrated in the spatial mapping of $dI/dV$ intensity at the Dirac point [Figs.~1(e) and 1(f)]. 
This observed phenomenon is attributed to the induced $sp^3$ orbital hybridization between chemisorbed atomic oxygen and graphene, which subsequently leads to the emergence of a  band gap around the Dirac point~\cite{dai2013DiffusionDesorptionOxygen,li2019ThermodynamicsKineticsOxygen}. 

As demonstrated above, we can tune the local electronic properties of graphene at the atomic scale by introducing atomic oxygen.
However, for device applications, a technique to spatially control the distribution of oxygen atoms is essential. 
As shown in Fig.~1(b), chemisorbed oxygen atoms are randomly distributed across the graphene surface.
To fabricate atom-scale electronic circuits with the desired size and structure, these randomly chemisorbed atoms need to be removed or relocated. 
Although thermal desorption has been previously reported as a method to remove oxygen atoms from graphene~\cite{Hossain2012a}, this method lacks the precise control needed for practical application.

\begin{figure} 
\includegraphics[width=8cm]{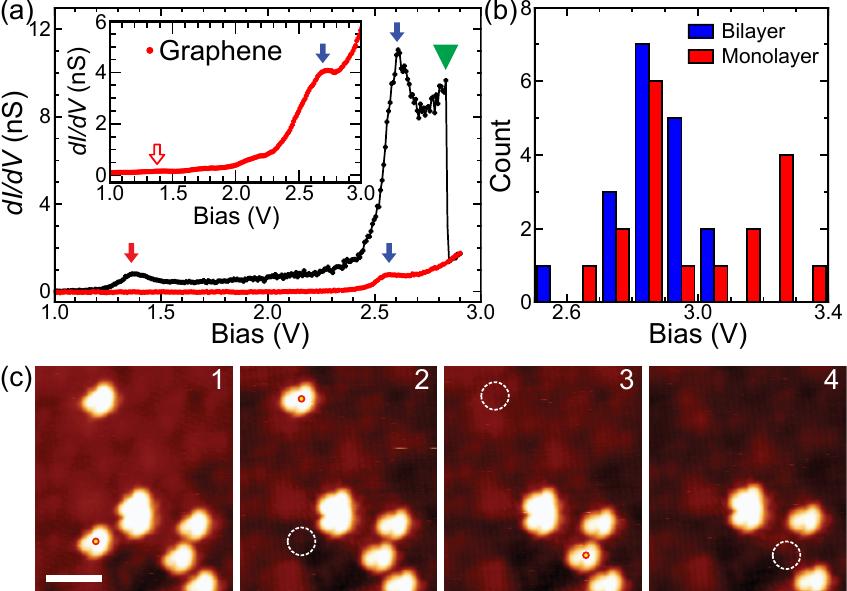}
\caption{\label{figure2}
(a) $dI/dV$ spectra recorded during the tip-induced desorption of atomic oxygen (set point: $V_{\rm s}$ = 1.0~V, $I_{\rm t}$ = 10~pA, $Z_{\rm offset}$ = 0~pm). 
The green triangle highlights the sharp decrease in tunneling current caused by oxygen desorption. 
The spectral features in the $dI/dV$ spectrum corresponding to atomic oxygen (indicated by a red arrow) and graphene on SiC (denoted by blue arrows) can be discerned at $+1.37$~V  and $+2.61$~V, respectively. 
The inset shows the $dI/dV$ spectrum obtained on the graphene surface under identical set point conditions.
(b) Histogram showing the bias voltage at which oxygen desorption occurred. 
(c) Sequential removal of chemisorbed atomic oxygen, demonstrating the feasibility of the controlled oxygen desorption on epitaxial graphene. 
The numbers indicate the order of removal. 
Small red circles represent the tip positions before oxygen desorption.
The scale bar measures 2~nm.}
\end{figure}

Chemisorbed oxygen atoms can be removed by injecting energetic electrons from the STM tip ($V_{\rm b}=+4.0$~V, $I_{\rm t}=1$~nA)~\cite{Hossain2012a}. 
However, the threshold bias or current has not been previously reported in detail.
To thoroughly investigate this, we employed a bias sweep method rather than applying a voltage pulse or scanning with high bias and current. 
Figure~2(a) shows the $dI/dV(V)$ curve, recorded while incrementally increasing the bias voltage from $V_{\rm b}=+1.0$~V ($I_{\rm t}=10$~pA) with the STM tip over an oxygen protrusion. 
During the forward bias sweep, we noted a sudden drop in differential conductance at $+2.83$~V, resulting in a differential conductance reduction by approximately a factor of 9. 
A subsequent topographic STM image affirmed this conductance drop resulted from the desorption of the oxygen atom during the bias sweep. 
Notably, the $dI/dV$ curve exhibits two distinct peaks during the forward sweep at $+1.37$~V and $+2.61$~V.
During the backward sweep, the peak at $+1.37$~V disappeared, while the peak at $+2.61$~V remained. 
This observation implies that the $+1.37$~V peak is associated with chemisorbed atomic oxygen, potentially attributable to the antibonding states from the $p_z$ orbitals of oxygen and adjacent carbon atoms~\cite{Suarez2011a}.
On the other hand, the $+2.61$~V peak seems linked with the graphene/SiC, as confirmed by the comparison curve from graphene [inset of Fig.~\ref{figure2}(a)]. 

We performed oxygen desorption by bias sweeps on more than 40 oxygen atoms on both monolayer and bilayer graphene.
Remarkably, we consistently observed similar $dI/dV$ curves, with the exception of the threshold bias. 
Figure~2(b) illustrates a histogram summarizing the desorption threshold bias on monolayer and bilayer graphene. 
The oxygen desorption phenomenon was not significantly affected by the tunneling current level, thereby ruling out mechanisms such as local heating induced by the tunneling current. 
The observed desorption threshold slightly exceeds the calculated adsorption energy of 2.2--2.4~eV~\cite{yi2017ContrastingDiffusionBehaviors,dai2013DiffusionDesorptionOxygen}, defined as $E_{\rm ad} = E_{\rm O/graphene}-E_{\rm graphene}-E_{\rm O}$.
This suggests that the removal of chemisorbed atomic oxygen from graphene may require more energy than its initial binding.
Interestingly, we observed a slightly different behavior between monolayer and bilayer graphene in terms of the oxygen desorption conditions categorized by the number of basal graphene layers [Fig.~\ref{figure2}(b)].
At present, no definitive explanation accounts for this observed difference, warranting further investigations to fully understand its underlying mechanism.

In the process of inelastic tunneling, the energy carried by tunneling electrons can be transferred to a molecule.
This may lead to the breaking of molecular chemical bonds if the transferred energy surpasses a dissociation barrier~\cite{Shen1995a,stipe1997,lauhon2000,hla2001,hla2003}.
By gradually increasing the energy of the tunneling electrons, it becomes possible to break the bond between atomic oxygen and carbon atoms when the energy transfer exceeds its dissociation barrier~\cite{yi2017ContrastingDiffusionBehaviors,dai2013DiffusionDesorptionOxygen}. 
This phenomenon of inelastic tunneling provides a plausible explanation for the broad distribution of desorption biases observed, as well as the presence of higher threshold biases. 

The controlled bias sweep technique can be used to detach individual oxygen atoms with high precision, as demonstrated in Fig.~2(c). 
Remarkably, neighboring atoms as close as $\sim2$~nm can remain unaffected by the desorption process, indicating a high degree of selectivity. 
This high selectivity is essential for the design and construction of atomic-scale electronic circuits, as it removes extra atoms while preserving the functionality of the device.

\begin{figure} 
\includegraphics[width=8cm]{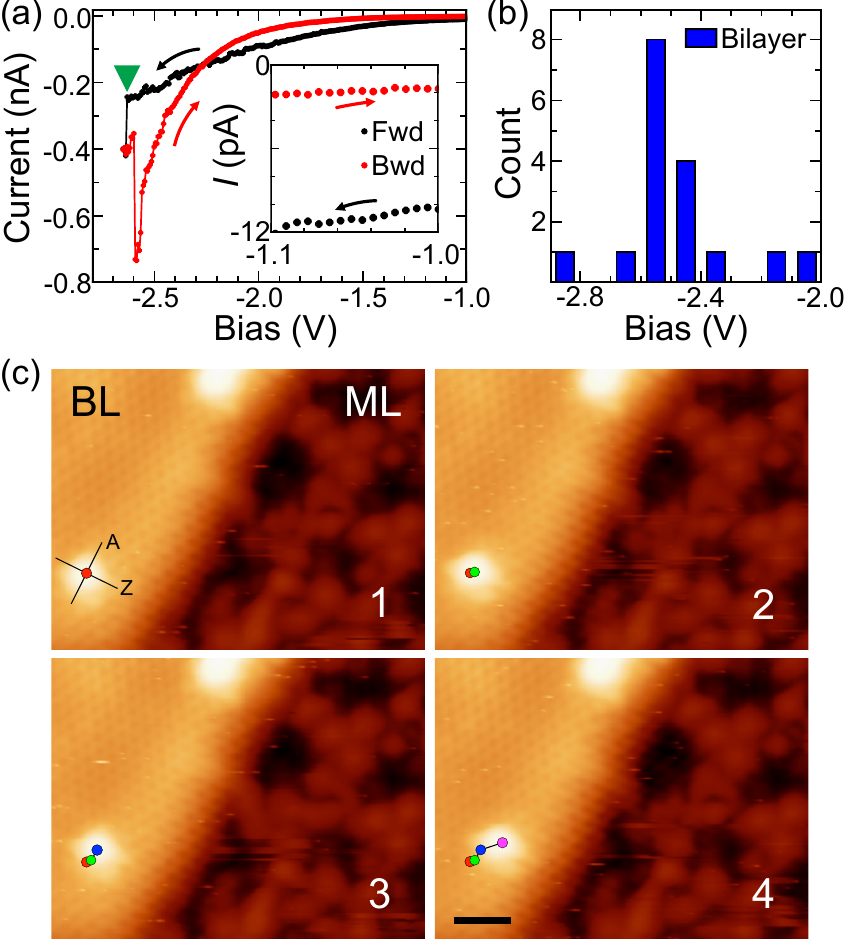}
\caption{\label{figure3}
(a) $I(V)$ curve recorded during the negative bias sweep inducing the lateral hopping of chemisorbed atomic oxygen on bilayer graphene (set point: $V_{\rm s}$ = $-1.0$~V, $I_{\rm t}$ = 10~pA, $Z_{\rm offset}$ = 0~pm). 
The green triangle highlights the abrupt increase in tunneling current due to the hopping event. 
(b) Histogram showing the bias at which oxygen hopping occurred.
(c) STM images showing sequential hopping near a bilayer-monolayer graphene step. 
Each colored circle represents the initial position of the oxygen atom prior to each hopping event.
The exact position of the oxygen atom was determined by obtaining line profiles along the zigzag (Z) and armchair (A) directions of graphene, as indicated.
The enumerated indicators signify the order of the hopping events. 
The scale bar is equivalent to 1~nm.
}
\end{figure}

When constructing atomic-scale artificial structures with an STM tip, it is often preferable to induce lateral movement of target atoms or molecules rather than controlling their desorption.
In this regard, we have made a significant discovery that the STM tip can control the lateral motion of chemisorbed oxygen specifically on bilayer graphene, achieved by applying a bias sweep method within negative bias ranges.
Figure~3(a) shows $I(V)$ curves as the sample bias is changed from $-1.0$~V to $-2.65$~V and vice versa, revealing a sudden increase in current.
Upon returning the bias to $-1$~V, the final tunneling current reduces to about $1/5$ of the initial value.
Compared to oxygen desorption, the final current exhibits a non-zero value, suggesting that the oxygen atom underneath the tip does not completely vanish.
Subsequent STM imaging validates the movement of the atom from its initial position during the bias sweep.
Analyzing the hopping condition as a histogram in Fig.~3(b), we determine an average threshold bias of $-2.48$~V, which has a smaller absolute value compared to the desorption condition. 
On rare occasions, we did observe oxygen desorption at a higher negative bias ($|V_{\rm b}|>3$~V), exclusively in cases where the targeted oxygen atoms did not move during the manipulating process. 
Compared to the positive bias sweep, the negative bias sweep was substantially less effective in removing atomic oxygen, indicating potentially different desorption mechanisms.

We have  demonstrated that lateral manipulation of atomic oxygen can be realized through negative bias sweeps.
Initially, we positioned the STM tip above a selected oxygen atom to be manipulated, applying negative bias sweeps until we detected an abrupt increase in tunneling current.
Subsequently, we employed an STM image to precisely locate the target oxygen atom. 
Through repetitive lateral manipulation, we were able to observe sequential hopping of the oxygen atom, as depicted in Fig.~3(c). 
Interestingly, the oxygen atom did not always relocate to the nearest bridge site but sometimes jumped to second or third neighboring positions, even including occasional longer-distance hopping. 

In-depth analysis of our experimental results reveals that directing the specific movement of oxygen atoms remains a considerable challenge. 
Given the limited number of atoms manipulated in our experiment, our observations may not be sufficient to draw a definitive conclusion.
However, the oxygen atom seems to roughly follow the step between bilayer and monolayer graphene [Fig.~3(c)].
This potential preference suggests more systematic experiments to uncover potential correlations between atom movement directions and factors like surface strain. 
To fully understand the underlying mechanism driving this lateral motion, further investigations are necessary.

In this study, we have investigated the properties of chemisorbed atomic oxygen on epitaxial graphene using STM. 
Our findings illustrate the band gap derived from the chemisorbed atomic oxygen near the Dirac point. 
Moreover, we have demonstrated that the oxygen atoms on graphene can be removed or relocated by applying gradually increasing bias voltages. 
Interestingly, the behavior of the oxygen atoms depends on the polarity of the bias voltage, allowing both removal and rearrangement using the same tip. 
While the precise mechanisms of desorption and hopping remain clear, this technique carries significant potential for tailoring the electronic properties of graphene using laterally predefined atomic oxygen. 
Such manipulation of oxygen atoms could pave the way for building atomic-scale electronic devices consisting only of oxidized graphene. 
Further understanding of the desorption and hopping mechanisms will improve the effectiveness of this technique in graphene-based device fabrication.

\begin{acknowledgments}
This work was supported by the Institute for Basic Science (Grant No. IBS-R014-D1) and the National Research Foundation of Korea (NRF) grant funded by the Korea government (MSIT) (NRF-2017R1A2B4007742 and 2022M3H4A1A04074153).
H.C.C. was supported by Samsung Electronics and Veteran researcher grant (No. 2019R1A2C2004259) managed by NRF.
\end{acknowledgments}

\subsection*{Data Availability Statement}
The data that support the findings of this study are available from the corresponding author upon reasonable request.


\begin{thebibliography}{47}
\bibitem [{sts()}]{sts}%
  \BibitemOpen
  \href@noop {} {}\bibinfo {note} {During the acquisition of the $dI/dV$
  spectra, the tunneling condition was established with parameters $V_{\rm s}$
  and $I_{\rm t}$. After the STM feedback was turned off, the tip's position
  was automatically adjusted by $Z_{\rm offset}$. Subsequently, the $dI/dV$
  signal was recorded as a function of the applied bias voltage.}\BibitemShut
  {Stop}%
\end{thebibliography}
%

\end{document}